\documentstyle[12pt,epsf]{article}
\begin{document}
\thispagestyle{empty}
\setcounter{page}{0}
\begin{flushright}
CERN-TH/96-310\\
RAL-96-092\\
October 1996\\
hep-ph/9611218
\end{flushright}
\vspace*{0.7cm}
\centerline{\large\bf Direct $J/\psi$ and $\psi'$ Polarization and 
Cross Sections}
\vspace*{0.1cm} 
\centerline{\large\bf at the Tevatron}
\vspace*{1.2cm}
\centerline{{\sc Martin Beneke$^1$} and {\sc Michael Kr\"amer$^2$}}
\bigskip
\centerline{\sl $^1$Theory Division, CERN,}
\centerline{\sl CH-1211 Geneva 23, Switzerland}
\vskip0.3truecm
\centerline{\sl $^2$ Rutherford Appleton Laboratory,}
\centerline{\sl Chilton, Didcot, OX11 0QX, England}

\vspace*{1cm}
\centerline{\bf Abstract}
\vspace*{0.5cm}
\noindent Transverse polarization of ${}^3 S_1$ charmonium states, produced 
directly in $p\bar{p}$ collisions at asymptotically large transverse
momentum $p_t$, has emerged as the most prominent test of color octet
contributions and spin symmetry in quarkonium production.  We present
predictions for the polar angle distribution at moderate values of
$p_t \sim 4 - 20\,$GeV, covered by the Tevatron Run I data.  We update
the fits of NRQCD matrix elements and discuss their theoretical
uncertainties. With our best fit values, no transverse polarization is
expected at $p_t\sim 5\,$ GeV, but the angular distribution is
predicted to change dramatically as $p_t$ increases to $20\,$ GeV.

\vspace*{0.5cm}

\vfill

\newpage

During the past three years, the phenomenology of quarkonium
production has undergone a phase of exciting developments. These
followed from the application of Non-Relativistic QCD (NRQCD)
\cite{BBL}, an effective theory that disentangles physics on the scale
of the heavy quark mass $m_Q$, relevant to the production of a heavy
quark pair, from physics on the scale of the bound state's binding
energy $m_Q v^2$, relevant to the formation of the quarkonium.
Inclusive quarkonium production is now viewed as a two-step process,
where the production of a heavy quark pair $Q\bar{Q}[n]$ in a certain
angular momentum and color state $n$ is described by a perturbatively
calculable short-distance cross section $d\hat{\sigma}_n$ and the
subsequent quarkonium formation is parametrized by a nonperturbative
matrix element $\langle {\cal O}_n^\psi \rangle$, which is subject to
the power counting rules of NRQCD.

For $J/\psi$ and $\psi'$ (collectively denoted by $\psi$) only the
color singlet $n={}^3\!S_1^{(1)}$ intermediate state contributes at
lowest order in the non-relativistic velocity expansion in $v^2$.  (We
use spectroscopic notation with the superscript in brackets denoting
the color state.) Three more nonperturbative parameters, related to
color octet configurations $n={}^3\!S_1^{(8)},\,{}^1\!S_0^{(8)},\,
{}^3\!P_J^{(8)}$ appear as $v^4$-corrections. Their importance for
$\psi$ production has been understood in \cite{BF}. Because gluons
couple easily to ${}^3\!S_1$ octet states, gluon fragmentation into
color octet quark pairs appears as the most plausible explanation of
the large direct $\psi$ production cross section observed at the
Tevatron \cite{CDF}. Subsequent to this observation many charmonium
production processes have been reconsidered to gauge the impact of
color octet contributions and to determine the corresponding
nonperturbative matrix elements. (For reviews and references see 
\cite{BEN3}.) Among the specific predictions of NRQCD,
transverse polarization of direct $\psi$ (i.e. not from $B$ and
$\chi_c$ decays) at large transverse momentum $p_t$ in $p\bar{p}$
collisions at the Tevatron has emerged as the most distinct and most
accessible signature \cite{CW,BEN1}. Transverse polarization also
discriminates NRQCD from other approaches, like the color evaporation
model which, in its most naive version, predicts the quarkonium to 
be produced unpolarized. At moderate $p_t$ the polarization 
signature provides an
important consistency check on the relative weight of the intermediate
color octet states inferred from the unpolarized differential cross
section.

Transverse polarization at large $p_t$ is a direct consequence of
gluon fragmentation as well as spin symmetry of the leading order
NRQCD effective lagrangian. When $p_t \gg 2 m_c$ the fragmenting gluon
is effectively on-shell and transverse. The $c\bar{c}$ pair in the
color octet ${}^3\!S_1$ state inherits the gluon's transverse
polarization and so does the $\psi$, because the emission of soft
gluons during hadronization does not flip the spin. Both, corrections
due to spin-symmetry breaking and higher order fragmentation
contributions have been estimated to be rather small \cite{BEN1}.
However, the color octet ${}^1\!S_0^{(8)}$ and ${}^3\!P_J^{(8)}$
production channels, which do not have a fragmentation interpretation
at leading order in the strong coupling $\alpha_s$, have large
short-distance coefficients \cite{CL} and can dominate the
$p_t$-distribution over a significant range of $p_t$ covered by the
present Tevatron data. The polarization yield from these subprocesses
has not so far been calculated.

In this Letter we report on the calculation of all subprocesses for
polarized direct $\psi$ (not from $B$ and $\chi_c$ decays) production
at leading order in $\alpha_s$.  We update the extraction of NRQCD
matrix elements from unpolarized cross sections first performed in
\cite{CL} and discuss their uncertainties.  The prediction for the
polar angle distribution includes the higher order fragmentation
corrections from \cite{BEN1} and covers the entire $p_t$-range
$p_t\sim 4-20\,$ GeV accessible with the Tevatron run I data. The
polarization analysis of Tevatron data
is under way \cite{VAIA}.\\

The differential cross section for $p\bar{p}\to \psi^{(\lambda)}(p_t)+X$ 
is given by
\begin{equation}
\frac{d\sigma}{d p_t}=\sum_{i,j}\int d x_1 d x_2\,f_{i/p}(x_1) 
f_{j/\bar{p}}(x_2)\, 
\sum_n \frac{d\hat{\sigma}_{ij}^{(\lambda)}
[n]}{d p_t}\,\langle{\cal O}^\psi_n\rangle,
\label{fact}
\end{equation}
where $f_{i/p}$ and $f_{j/\bar{p}}$ denote the parton densities and
$\lambda$ specifies the helicity state.  At non-vanishing transverse
momentum, the leading partonic subprocesses $i+j\to c\bar{c}[n]+k$
occur at order $\alpha_s^3$. To obtain the short-distance cross
section for a given $c\bar{c}$ state $n$, we expand, in the rest frame
of the heavy quark pair, the partonic amplitude in the relative
momentum of the heavy quarks and decompose the amplitude in spin and
color. The amplitude squared for $i+j\to\psi^{(\lambda)}+X$ can now be
written as
\begin{equation}
{\cal M} = \sum_n {\cal M}_{n;kl\ldots}^{(\lambda)}\,
\langle{\cal O}^{\psi^{(\lambda)}}_
{n;kl\ldots}\rangle,
\end{equation}
where the sum is ordered as an expansion in $v^2$. The cartesian tensors 
\begin{equation}
\langle{\cal O}^{\psi^{(\lambda)}}_
{n;kl\ldots}\rangle = 
\sum_X\,\langle 0|\,\chi^\dagger {\cal \kappa}_{n;k\ldots}\psi\,
|\psi^{(\lambda)}X
\rangle\langle \psi^{(\lambda)}X|\,\psi^\dagger {\cal \kappa}^\prime_
{n;l\ldots}
\chi\,
|0\rangle
\end{equation}
describe the nonperturbative transition of the $c\bar{c}$ pair into 
a polarized quarkonium $\psi$ and any number of light hadrons $X$. 
Up to $v^6$-corrections to the color singlet contribution,  
Lorentz decomposition of these tensors, together with spin symmetry, 
allows us to reduce the matrix elements to four nonperturbative 
parameters 
$\langle {\cal O}_1^{\psi}({}^3S_1)\rangle$, 
$\langle {\cal O}_8^{\psi}({}^3S_1)\rangle$, 
$\langle {\cal O}_8^{\psi}({}^1S_0)\rangle$, 
$\langle {\cal O}_8^{\psi}({}^3P_0)\rangle$. They are 
defined in \cite{BBL} and include a sum over polarizations $\lambda$. 
We do not consider color singlet operators such as 
${\cal P}_1^\psi(^3\!S_1)$, in the notation of \cite{BBL}, whose 
contributions are expected to be irrelevant.

For unpolarized production, the calculational procedure sketched above 
leads to results identical to those obtained with the standard amplitude 
projection method \cite{GUB80}. The polarized production cross section, 
however, requires special care as interference contributions from 
intermediate $c\bar{c}$ pairs in different angular momentum states 
$J J_z$ no longer cancel \cite{BEN1,BEN2,BC96}. In this case, the 
relevant projections are unambiguously specified by 
rotational invariance and spin symmetry implemented in the decomposition of 
$\langle{\cal O}^{\psi^{(\lambda)}}_{n;kl\ldots}\rangle$ and can be found in 
\cite{BEN1}. For the $S=0$, $L=0$ intermediate state, 
each helicity state contributes one third of the 
unpolarized cross section.  For the $P$-wave intermediate state, 
one must first sum over all orbital angular momentum states $L_z$ and then 
project with the polarization vector $\epsilon(\lambda)$ of the 
quarkonium. The short-distance coefficients that enter 
$d\hat{\sigma}_{ij}^{(\lambda)}[n]$ can now be written as 
\begin{equation}
\label{res}
A_{ij}[n] + 
B_{ij}[n]\,(\epsilon(\lambda)\cdot k_1)^2 + 
C_{ij}[n]\,(\epsilon(\lambda)\cdot k_2)^2 + 
D_{ij}[n]\,(\epsilon(\lambda)\cdot k_1)\, (\epsilon(\lambda)\cdot k_2),
\end{equation}
where $k_1$ and $k_2$ are the momenta of the initial state partons and
$ij=gg,gq,g\bar{q},q\bar{q}$.  Detailed expressions for the
coefficients $A,\ldots,D$ will be presented elsewhere. Summing over
all polarizations $\lambda$, we find agreement with the
unpolarized cross sections computed in \cite{CL}.\\

\begin{figure}[t]
   \vspace{-2.3cm}
   \epsfysize=14cm
   \epsfxsize=10cm
   \centerline{\epsffile{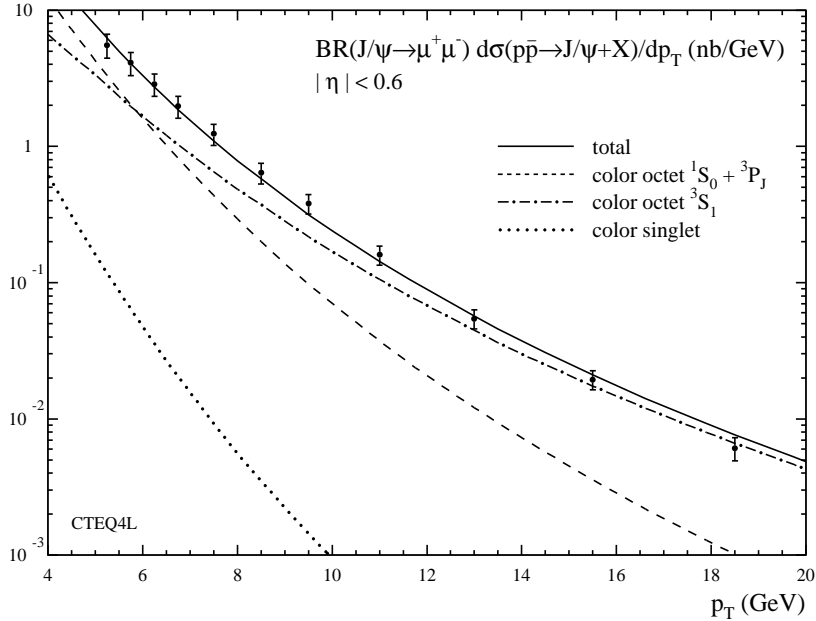}}
   \vspace*{-3cm}
\caption[dummy]{\label{jpsi} Fit of color octet contributions to direct 
  $J/\psi$ production data from CDF ($\sqrt{s}=1.8\,$TeV,
  pseudorapidity cut $|\eta|<0.6$).  The theoretical curves are
  obtained with CTEQ4L parton distribution functions, the
  corresponding $\Lambda_4=235\,$MeV, factorization scale
  $\mu=(p_t^2+4 m_c^2)^{1/2}$ and $m_c=1.5\,$GeV. The fitted color
  octet matrix elements are given as the central values in
  Tab~\ref{tab1}. }
\end{figure}
\begin{figure}[t]
   \vspace{-2.3cm}
   \epsfysize=14cm
   \epsfxsize=10cm
   \centerline{\epsffile{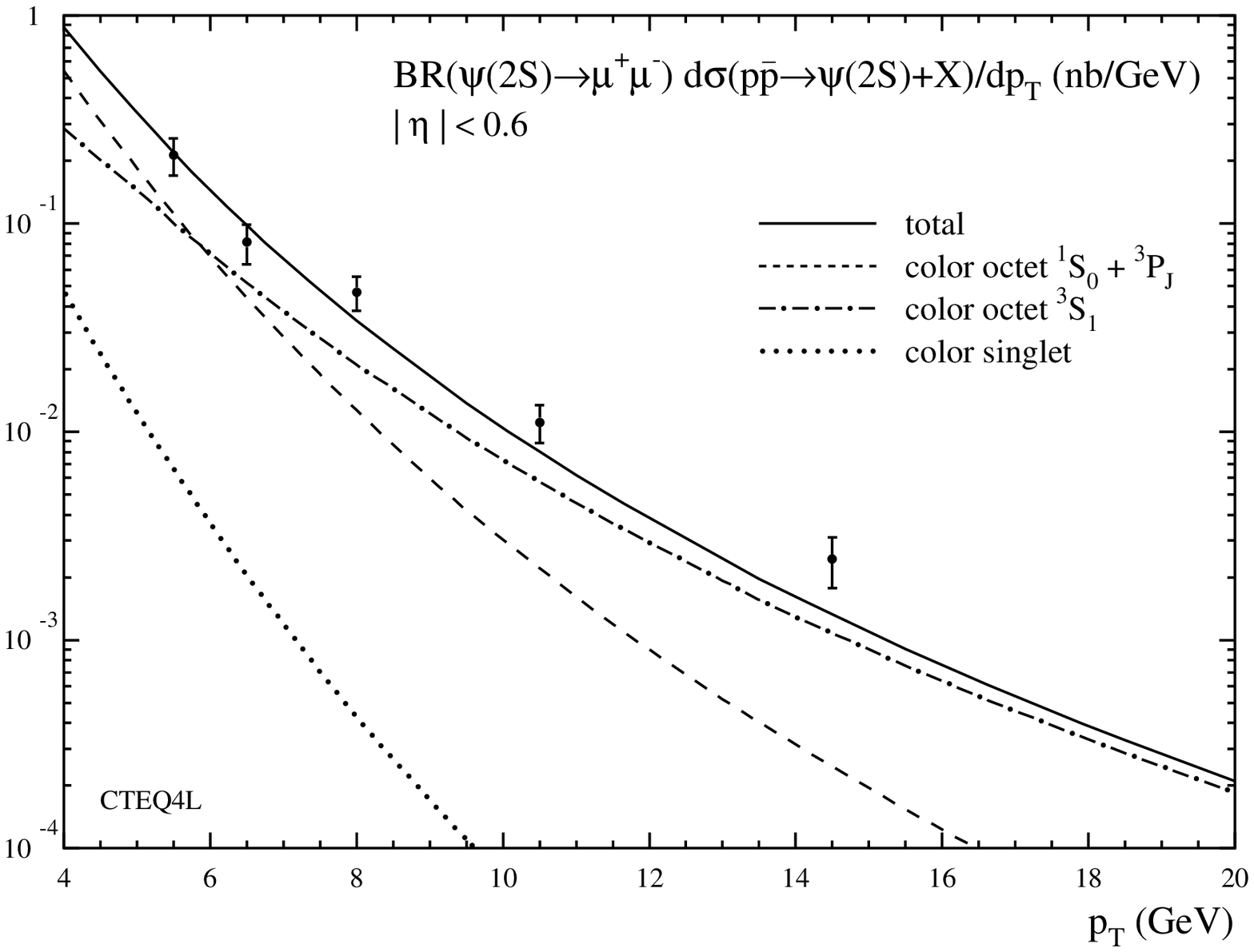}}
   \vspace*{-3cm}
\caption{\label{psii} Same as Fig.~\ref{jpsi} for prompt $\psi'$ production.}
\end{figure}

The nonperturbative matrix elements $\langle {\cal O}_n^\psi \rangle$
that enter the color octet contributions can be adjusted to fit the
experimental data as shown in Figs.~\ref{jpsi} and \ref{psii}, where
we compare the transverse momentum distributions for unpolarized
direct $J/\psi$ and $\psi'$ production with the most recent CDF data
\cite{data} ($\sqrt{s}=1.8\,$TeV).\footnote{For direct $J/\psi$
  production, indirect contributions from $\chi$ decays are removed,
  but those from $\psi'$ decays are included in both the data and
  theoretical prediction. The indirect $\psi'$ contribution amounts to
  about $11\%$ \cite{data}. In accounting for this contribution,
  we assumed that it is independent of $p_t$.} At large $p_t$, 
the cross section is
dominated by gluon fragmentation into color octet $^3\!S_1$ charm
pairs \cite{BF,CAC95}.  We have resummed the leading logarithms
$(\alpha_s\ln p_t^2/(2 m_c)^2)^n$ in this channel by solving the
Altarelli-Parisi evolution equation for the corresponding
fragmentation function. Evolution decreases the short-distance cross
section and enhances the fitted value of $\langle {\cal
  O}_8^{\psi}({}^3S_1)\rangle$. The color octet ${}^1S_0$ and
${}^3P_J$ channels are significant in the region $p_t\; \rlap{\lower
  3.5 pt \hbox{$\mathchar \sim$}} \raise 1pt \hbox {$<$}\; 10$~GeV,
but fall as $d\hat{\sigma}/dp_t^2\sim1/p_t^6$ and become negligible
for the unpolarized cross section at large $p_t$. Because the
${}^1S_0^{(8)}$ and ${}^3P_J^{(8)}$ short-distance cross sections have
a similar $p_t$ dependence, the transverse momentum distribution is
sensitive only to a combination
\begin{equation}
\label{k}
M_k^\psi(^1\!S_0^{(8)},^3\!P_0^{(8)})\equiv
\langle {\cal O}_8^{\psi}({}^1S_0)\rangle + \frac{k}{m_c^2}\,
\langle {\cal O}_8^{\psi}({}^3P_0)\rangle.
\end{equation}
As $p_t$ increases from $4$ to $20\,$GeV, $k$ varies from $3.9$ to
$3.0$ for $m_c=1.5\,$GeV.  Since the $^1\!S_0^{(8)}$ and
$^3\!P_J^{(8)}$ channels are significant only at low $p_t$, we choose
$k=3.5$. [Note that $k=3$ was chosen in \cite{CL}.] The theoretical
prediction then depends on two phenomenological parameters.

As in \cite{CL}, we fit these two parameters to the
$p_t$-distribution, but attempt to estimate, at least partially, the
theoretical uncertainties by varying the parton distribution function
(PDF) set and renormalization (factorization) scale between $1/2
\sqrt{p_t^2+4 m_c^2}$ and $2 \sqrt{p_t^2+4 m_c^2}$.  The result is
shown in Tab.~\ref{tab1} for three PDF sets \cite{pdf}, where the
first error is statistical only and the second reflects the scale
variation.  In all cases the coupling $\alpha_s(\mu)$ is evaluated
according to the $\Lambda_4$ specified by the chosen PDF set.  In case
of MRS(R2) this implies that $\alpha_s(\mu)$ is evolved with two-loop
accuracy.

While the value of $\langle {\cal O}_8^{\psi}({}^3S_1)\rangle$ is
relatively stable with respect to the uncertainty related to the PDF,
the fit of $M_{3.5}^\psi(^1\!S_0^{(8)},^3\!P_0^{(8)})$ depends
sensitively on all effects that modify the slope of the
$p_t$-distribution, 
such as the small-$x$ behaviour of the gluon distribution or the
evolution of the strong coupling.  A steeper gluon distribution
increases the slope of the $p_t$-distribution and leads to smaller
values of $M_{3.5}^\psi(^1\!S_0^{(8)},^3\!P_0^{(8)})$. In addition,
next-to-leading contributions in $\alpha_s$ and also systematic
effects inherent to NRQCD, such as the inaccurate treatment of energy
conservation in the hadronization of the color octet quark pairs,
modify the shape of the $p_t$-distribution, but are not reflected in
the theoretical errors quoted in Tab.~\ref{tab1}. In view of this, the
values of $M_{3.5}^\psi(^1\!S_0^{(8)},^3\!P_0^{(8)})$ should be
considered with caution.

\begin{table}[t]
\renewcommand{\arraystretch}{1.25}
$$
\begin{array}{cccc}
\hline\hline
& \mbox{CTEQ4L} & \mbox{GRV (1994) LO} & \mbox{MRS(R2)} \\ 
\hline
\langle {\cal O}_8^{J/\psi}({}^3S_1)\rangle & 
1.06\pm0.14_{-0.59}^{+1.05} & 1.12\pm0.14_{-0.56}^{+0.99} & 
1.40\pm0.22_{-0.79}^{+1.35} \\
M_{3.5}^{J/\psi}(^1\!S_0^{(8)},^3\!P_0^{(8)}) & 
4.38\pm1.15_{-0.74}^{+1.52} & 3.90\pm1.14_{-1.07}^{+1.46} & 
10.9\pm2.07_{-1.26}^{+2.79} \\
\langle {\cal O}_8^{\psi'}({}^3S_1)\rangle & 
0.44\pm0.08_{-0.24}^{+0.43} & 0.46\pm0.08_{-0.23}^{+0.41} & 
0.56\pm0.11_{-0.32}^{+0.54} \\
M_{3.5}^{\psi'}(^1\!S_0^{(8)},^3\!P_0^{(8)}) & 
1.80\pm0.56_{-0.30}^{+0.62} & 1.60\pm0.51_{-0.44}^{+0.60} & 
4.36\pm0.96_{-0.50}^{+1.11}  \\ 
\hline\hline
\end{array}
$$
\caption{\label{tab1}
NRQCD matrix elements in $10^{-2}\,$GeV$^3$. First error statistical, 
second error due to variation of scale. Ratio of $\psi'$ to $J/\psi$ 
fixed.}
\end{table}

For $\psi'$ production, an unconstrained fit of $\langle {\cal
  O}_8^{\psi'}({}^3S_1)\rangle$ and
$M_{3.5}^{\psi'}(^1\!S_0^{(8)},^3\!P_0^{(8)})$ leads to a value of the
second parameter compatible with zero (with large error), or slightly
negative for GRV and CTEQ4L PDFs.  We conclude, that the present data
does not allow a meaningful fit of this parameter. We therefore
performed a combined fit of $J/\psi$ and $\psi'$ data under the
(arguably plausible) assumption that the ratio
$M_{3.5}^{\psi}(^1\!S_0^{(8)},^3\!P_0^{(8)})/ \langle {\cal
  O}_8^{\psi}({}^3S_1)\rangle$ is the same for $J/\psi$ and $ \psi'$.
The corresponding results for $\psi'$ are also shown in
Tab.~\ref{tab1}.\\

We next consider charmonium polarization in the ($s$-channel) 
helicity frame (`recoil frame'). In this frame the polarization axis 
in the $\psi$ rest frame is defined as the direction of the $\psi$ 
three-momentum in the hadronic cms frame. Let $P=p_p+p_{\bar{p}}$ be 
the sum of the initial hadron four-momenta and $Q$ the $\psi$ four-momentum. 
In the $\psi$ rest frame $\epsilon_L=\epsilon(\lambda=0)=(0,-\vec{P}/|
\vec{P}|)$. The covariant expression reads
\begin{equation}
\epsilon_L(Q)_\mu = \frac{P\cdot Q}{\sqrt{(P\cdot Q)^2-M^2 s}}\,
\left(\frac{Q_\mu}{M}-\frac{M}{P\cdot Q}\,P_\mu\right),
\end{equation}
where $M$ is the quarkonium mass, taken to be $2 m_c$, and $\sqrt{s}$
the hadronic cms energy.  We then obtain the longitudinal polarization
fraction $\xi=d\sigma^{(\lambda=0)}/\sum_\lambda d\sigma^{(\lambda)}$.
The polar angular distribution in $\psi\to l^+ l^-$ decay is given by
\begin{equation}
\label{angular}
\frac{d\Gamma}{d\cos\theta}\propto 1+\alpha\,\cos^2\theta,
\end{equation}
with $\alpha=(1-3\xi)/(1+\xi)$ and $\theta$ the angle between the
lepton three-momentum in the $\psi$ rest frame and the polarization
axis, the $\psi$ direction in the hadronic cms frame (lab frame).
Note that in this frame scalar products like $\epsilon_L(Q)\cdot k_1$
[see Eq.~(\ref{res})] depend explicitly on parton momentum fractions
$x_i, x_j$ and can not be expressed in terms of partonic Mandelstam
variables alone. For this reason, the partonic helicity amplitudes for
the $^3\!S_1^{(8)}$ channel computed in \cite{CL} can not be used for
the present purpose.

\begin{table}[t]
$$
\begin{array}{ccccc}
\hline\hline
p_t/\mbox{GeV} & a_L & b & c & c_L \\ 
\hline
 4 & 0.144 & 0.457 & 1.81  & 0.896 \\
 8 & 0.059 & 0.147 & 0.480 & 0.191 \\
12 & 0.031 & 0.074 & 0.230 & 0.085 \\ 
16 & 0.019 & 0.047 & 0.144 & 0.051 \\
20 & 0.013 & 0.033 & 0.099 & 0.034 \\
24 & 0.010 & 0.024 & 0.073 & 0.025 \\
\hline\hline
\end{array}
$$
\caption{\label{tab2}
Coefficients entering the longitudinal polarization fraction. Parameter 
specifications and pseudorapidity cut as for Fig.~\ref{jpsi}.}
\end{table}

\begin{figure}[t]
   \vspace{-2.3cm}
   \epsfysize=14cm
   \epsfxsize=10cm
   \centerline{\epsffile{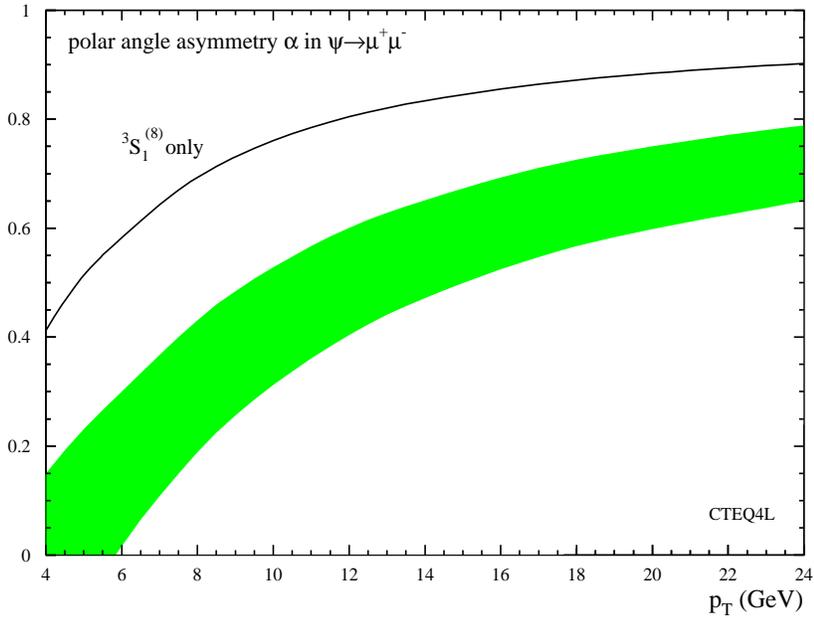}}
   \vspace*{-3cm}
\caption[dummy]{\label{alpha} $\alpha$ as a function of $p_t$. Parameter 
specifications as for Fig.~\ref{jpsi}. The band is obtained 
from Eq.~(\ref{xil}) and the ${\cal O}(\alpha_s^4)$ fragmentation 
contributions from \cite{BEN1} as described in the text. The solid 
curve is obtained with $M_{3.5}^{\psi}(^1\!S_0^{(8)},^3\!P_0^{(8)})=0$ 
and shows the importance of this parameter for polarization. The band 
does not include theoretical uncertainties in the short-distance 
coefficients, Tab.~\ref{tab2}.}
\end{figure}
l
In the large-$p_t$ limit, the $^3\!S_1^{(8)}$ production channel 
yields transverse polarization, while the $^1\!S_0^{(8)}$ and 
$^3\!P_J^{(8)}$ channels both yield unpolarized quarkonia in this 
limit. However, for the longitudinally polarized cross section 
$k$, defined in Eq.~(\ref{k}), varies from $5.9$ to $3.1$ as 
$p_t$ increases from $4$ to $20\,$GeV, so that in the important 
region of low $p_t$ a combination of 
$\langle {\cal O}_8^{\psi}({}^1S_0)\rangle$ and  
$\langle {\cal O}_8^{\psi}({}^3P_0)\rangle$ different from 
$M_{3.5}^{\psi}(^1\!S_0^{(8)},^3\!P_0^{(8)})$ is required. 
We represent the longitudinal polarization fraction $\xi$ as 
\begin{equation}
\label{xil}
\xi = \frac{a_L \langle {\cal O}_8^{\psi}({}^3S_1)\rangle + 
b_L \langle {\cal O}_8^{\psi}({}^1S_0)\rangle + 
c_L \langle {\cal O}_8^{\psi}({}^3P_0)\rangle/m_c^2}
{\langle {\cal O}_8^{\psi}({}^3S_1)\rangle + 
b \langle {\cal O}_8^{\psi}({}^1S_0)\rangle + 
c \langle {\cal O}_8^{\psi}({}^3P_0)\rangle/m_c^2}.
\end{equation}
The color singlet contribution is negligible.  The coefficients
$a_L,c_L,b,c$ are shown in Tab.~\ref{tab2} as function of $p_T$. Note
that $b_L=b/3$ exactly. Fig.~\ref{alpha} displays $\alpha$, defined in
Eq.~(\ref{angular}) as function of $p_t$. The shaded band is obtained
as a combination of the 
uncertainty (statistical only) in the extraction of the NRQCD
matrix elements [Tab.~\ref{tab1}] and the limiting cases that either
$\langle {\cal O}_8^{\psi}({}^1S_0)\rangle$ or $\langle {\cal
  O}_8^{\psi}({}^3P_0)\rangle$ is set to zero in the combination
$M_{3.5}^{\psi}(^1\!S_0^{(8)},^3\!P_0^{(8)})$.  Note that $\alpha$ as
a function of $p_t$ is identical for $J/\psi$ and $\psi'$ production,
because the ratios of NRQCD matrix elements for $J/\psi$ and $\psi'$
have been fixed. Thus, in comparing Fig.~\ref{alpha} with direct
$J/\psi$ polarization data, $\psi'$ feed-down must be removed in
addition to $\chi$ feed-down. At low $p_t\sim 5\,$GeV, the theoretical
prediction is compatible with unpolarized quarkonia. As $p_t$
increases, the angular distribution becomes more and more anisotropic.
At $p_t\sim 15\,$GeV, the fragmentation contributions of order
$\alpha_s^4$ computed in \cite{BEN1} become equally important as the
$\alpha_s^3$ mechanisms computed here.\footnote{In Eq.~(13) of
  \cite{BEN1} the factor 864 should be replaced by 96. The values of
  $r_3$ in Tab.~1 of \cite{BEN1} must be multiplied by three.} This
second source of longitudinal polarization
is included in the curves in Fig.~\ref{alpha}.\\

In conclusion, we have evaluated the dominant sources of direct
charmonium depolarization in $p\bar{p}$ collisions at moderate
transverse momentum. In the lower $p_t$ range about $5\,$GeV, which,
due to limitation in statistics, will be most relevant in the present
Tevatron analysis, an essentially flat angular distribution is
predicted, unless the $\langle {\cal O}_8^{\psi}({}^1S_0)\rangle$ and
$\langle {\cal O}_8^{\psi}({}^3P_0)\rangle$ matrix elements are
significantly smaller than their best fit values, obtained from the
$p_t$ distribution of the unpolarized cross section. With increasing
$p_t$, the angular distribution is predicted to change rapidly due to
the onset of substantial transverse polarization. Observation of a
pattern as in Fig.~\ref{alpha} would provide support
for the color octet charmonium production picture. \\

While this paper was in writing, charmonium polarization at moderate
$p_t$ has also been addressed by Leibovich \cite{LEI}.  Unfortunately,
contrary to what is stated in the paper, he defines the polarization
axis as the $\psi$ direction in the partonic (rather than hadronic)
cms frame. His results for the polar angle distribution can therefore
not be used for comparison with Tevatron data.\\

We thank M. Mangano and V. Papadimitriou for discussions 
and for providing us with CDF data.

\vfill\eject

\end{document}